\documentclass[aps,prx,reprint,longbibliography]{revtex4-1}
\usepackage{amsmath,amssymb,bm,graphicx,subfigure,enumerate,latexsym,comment,dcolumn}
\usepackage[dvips]{epsfig}

\begin{document}

\title{Synchronized molecular dynamics simulation via macroscopic heat and momentum transfer: an application to polymer lubrication}
\author{
Shugo Yasuda$^1$
\footnote{Electronic mail: yasuda@sim.u-hyogo.ac.jp}
and
Ryoichi Yamamoto$^2$
\footnote{Electronic mail: ryoichi@cheme.kyoto-u.ac.jp}
}
\affiliation{
$^1$Graduate School of Simulation Studies, University of Hyogo,
Kobe 650-0047, Japan,\\
$^2$Department of Chemical Engineering,
Kyoto University, Kyoto 615-8510, Japan}
%\date{\today}

\begin{abstract}
A synchronized molecular dynamics simulation via macroscopic heat and momentum transfer is proposed to model the non-isothermal flow behaviors of complex fluids.
In this method, the molecular dynamics simulations are assigned to small fluid elements to calculate the local stresses and temperatures and are synchronized at certain time intervals to satisfy the macroscopic heat- and momentum-transport equations.
This method is applied to the lubrication of a polymeric liquid composed of short chains of ten beads between parallel plates.
The rheological properties and conformation of the polymer chains coupled with local viscous heating are investigated with a non-dimensional parameter, the Nahme-Griffith number, which is defined as the ratio of the viscous heating to the thermal conduction at the characteristic temperature required to sufficiently change the viscosity.
The present simulation demonstrates that strong shear thinning and a transitional behavior of the conformation of the polymer chains are exhibited with a rapid temperature rise when the Nahme-Griffith number exceeds unity.
The results also clarify that the reentrant transition of the linear stress-optical relation occurs for large shear stresses due to the coupling of the conformation of polymer chains and due to the heat generation under shear flows.
\end{abstract}


%%%%%%%%%%%%%%%%%%%%%%%

\maketitle

\section{Introduction}
To predict the transport phenomena of complex fluids caused by the coupled heat and momentum transfer processes is challenging from both a scientific and engineering point of view. Molecular dynamics (MD) simulations are often used to predict material properties (e.g., the rheological, thermal, and electrical properties), in which the simulation is performed for a very small piece of the material under a certain ideal environment.\cite{book:89AT,book:08EM} However, in actual engineering and biological systems, the macroscopic features of complex fluids are highly affected by the spatial heterogeneity caused by the macroscopic transport phenomenon involved in the boundary conditions. A typical example is the generation of heat in lubrication systems.\cite{book:87BAH} To predict such complicated behavior in complex fluids, the entire system, including the boundary conditions, must be considered on the basis of an appropriate molecular model. In principle, full MD simulations of the entire system can meet these requirements. However, it is difficult in practice to perform full MD simulations on the macroscopic scale, which is common in actual engineering systems and is far beyond the molecular size. Multiscale modeling is a promising candidate to address this type of problem. 

The multiscale simulation for the flow behaviors of complex fluids was first advanced in the CONNFFESSIT approach for polymeric liquids by Laso and \"Ottinger\cite{art:93LO,art:95FLO,art:97LPO}, where the local stress in the fluid solver is calculated using the microscopic simulation instead of using any constitutive relations. The GENERIC approach is also presented for the non-isothermal polymeric flows by making important corrections and clarifications to the CONNFFESSIT scheme.\cite{art:99DEO} The strategy exploited in the CONNFFESSIT approach is also introduced into the heterogeneous multiscale modeling (HMM), which was proposed by E and Enquist,\cite{art:03EE} as a general methodology for the efficient numerical computation of problems with multiscale characteristics. HMM has been applied to various problems, such as the simple polymeric flow\cite{art:05RE}, coarsening of a binary polymer blend\cite{art:11MD} and the channel flow of a simple Lennard-Jones liquid\cite{art:13BLR}. The equation-free multiscale computation proposed by Kevrekids et al. is also based on a similar idea and has been applied to various problems.\cite{art:03KGHKRT, art:09KS} De et al. proposed the scale-bridging method, which can correctly reproduce the memory effect of a polymeric liquid, and demonstrated the non-linear viscoelastic behavior of a polymeric liquid in slab and cylindrical geometries.\cite{art:06DFSKK, art:13D} The multiscale simulation for polymeric flows with the advection of memory in two and three dimensions was developed by Murashima and Taniguchi.\cite{MT2010,MT2011,MT2012} We have also developed a multiscale simulation of MD and computational fluid dynamics (CFD). The multiscale method was first developed for simple fluids\cite{art:08YY} and subsequently extended to polymeric liquids with the memory effect.\cite{art:09YY,art:10YY,art:11YY,art:13MYTY}

However, a multiscale simulation for the coupled heat and momentum transfer of complex fluids has yet to be proposed. For complex fluids, it is usually difficult to describe the heat generation coupled with the momentum transport using only macroscopic quantities.
The spatial variation in the temperature also becomes notable at the macroscopic scale due to local viscous heating under shear flow. Thus, multiscale modeling is important for the coupled heat and momentum transfer of complex fluids. In the present paper, we propose a multiscale simulation, termed the synchronized molecular dynamics simulation (SMD), for the coupled heat and momentum transfer in complex fluids by extending the multiscale simulation for momentum transport and apply it to the polymer lubrication. Using this method, we investigate the rheological properties and conformation of polymer chains for the thermohydrodynamic lubrication of a polymeric liquid composed of short chains in a gap between parallel plates, in which the width of the gap is sufficiently large compared to the characteristic length of the flow behaviors, e.g., the length of the viscous boundary layer, such that the macroscopic quantities, e.g., the velocity, stress and temperature, become spatially heterogeneous. 

The full MD simulations for confined short-chain molecules in slab geometry were previously studied in Refs. \cite{art:99JAT} and \cite{art:01KPY}. The viscous heating of simple liquids in the same geometry was also studied using a full MD simulation in Refs. \cite{art:97KPY} and \cite{art:10KBC}. However, these results are exclusively for the molecularly thin films, i.e., the slab width is approximately ten times the molecular size, whereas in the present study, a width of over a thousand times the molecular size is considered in the SMD simulation.

In the following, we first describe the problem considered in the present paper. The simulation method is explained after the presentation of the problem. The SMD simulation of polymer lubrication is performed, and the results are discussed; these results are mainly the rheological properties and the coupling of the conformations of the polymer chains with heat generation under shear flows. We also present a critical analysis of the conceptual and technical issues of the SMD method. Finally, a short summary and a perspective on the future of SMD are given.

\begin{figure*}[tbp]
\begin{center}
\includegraphics[scale=1]{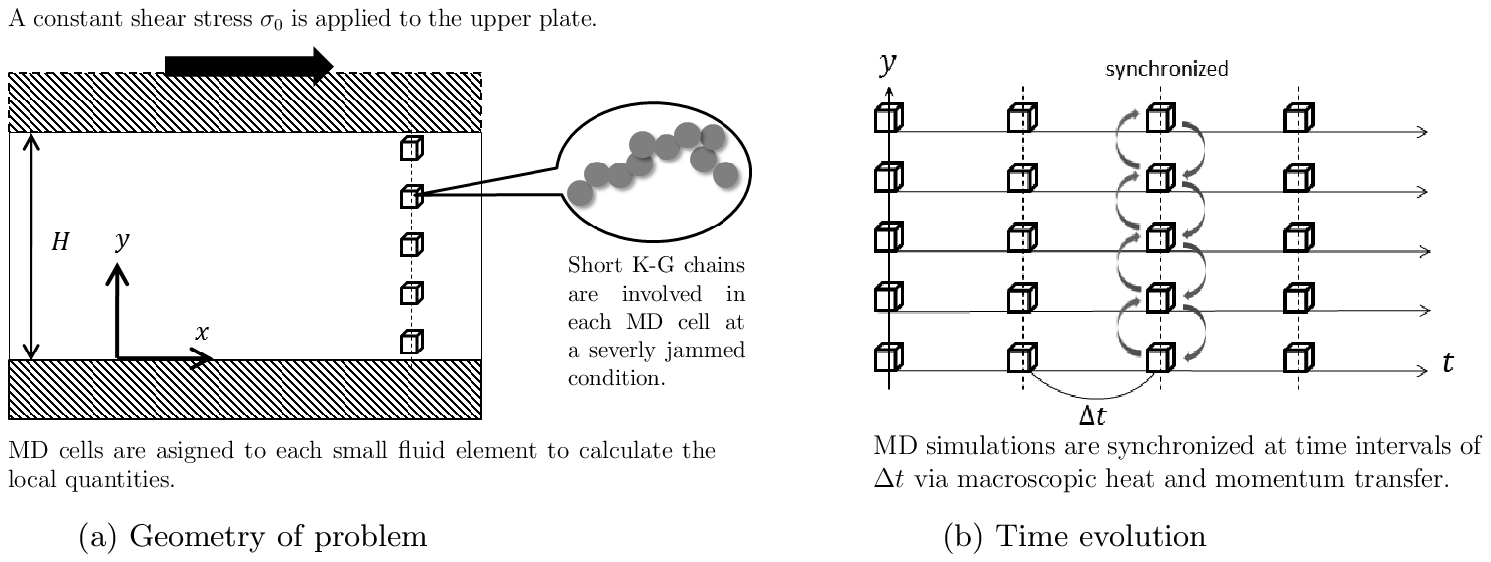}
\caption{%
The schematic of the geometry of the problem (a) and the time evolution (b).
}\label{fig_problem}
\end{center}
\end{figure*}
\section{Problem}
We consider a polymeric liquid contained in a gap of width $H$ between parallel plates with a constant temperature $T_0$ (see Fig. \ref{fig_problem}(a)). The polymeric liquid is composed of short Kremer-Grest chains\cite{art:90KG} of ten beads, in which all of the bead particles interact with a truncated Lennard-Jones potential defined by
\begin{equation}
U_{\rm LJ}(r)=\left\{
\begin{array}{l r}
4\epsilon\left[
\left(\frac{\sigma}{r}\right)^{12}
-\left(\frac{\sigma}{r}\right)^{6}
\right]
+\epsilon,
&\quad (r\le 2^{1/6}\sigma),\\
0,
&\quad (r\ge 2^{1/6}\sigma),
\end{array}
\right.
\end{equation}
and consecutive beads on each chain are connected by an anharmonic spring potential,
\begin{equation}
U_{\rm F}(r)=-\frac{1}{2}k_c R_0^2 \ln
\left[
1-\left(\frac{r}{R_0}\right)^2
\right],
\end{equation}
with $k_c$=30$\epsilon/\sigma^2$ and $R_0$=$1.5\sigma$. The polymeric liquid is in a quiescent state with a uniform density $\rho_0$ and a uniform temperature $T_0$ before a time $t=0$. Hereafter, the $y$-axis is perpendicular to the parallel plates, and the boundaries between the upper and lower plates and the polymeric liquid are located at $y=H$ and 0, respectively. 
The upper plate starts to move in the x direction with a constant shear stress $\sigma_0$ at a time $t$=0, while the lower plate is at rest.

The macroscopic behavior of the polymeric liquid is described by the following transport equations:
\begin{subequations}\label{eq_macro}
\begin{align}
\rho_0\frac{\partial v_x}{\partial t}&=\frac{\partial \sigma_{xy}}{\partial y},
\label{eq_flow}\\
\rho_0\frac{\partial e}{\partial t}&=\sigma_{xy}\dot\gamma - \lambda \frac{\partial^2 T}{\partial y^2},
\label{eq_ene}
\end{align}
\end{subequations}
where $v_\alpha$ is the velocity, $\sigma_{\alpha\beta}$ is the stress tensor, $e$ is the internal energy per unit mass, and $\dot \gamma$ is the shear rate, i.e., $\dot \gamma=\partial v_x/\partial y$. Hereafter, the subscripts $\alpha$, $\beta$, and $\gamma$ represent the index in Cartesian coordinates, i.e., \{$\alpha$,$\beta$,$\gamma$\}$\in$\{$x$,$y$,$z$\}. Here, we assume that the macroscopic quantities are uniform in the $x$ and $z$ directions, $\partial /\partial x$=$\partial /\partial z$=0, and the density of the polymeric liquid is constant.
Fourier's law for a heat flux with a constant and uniform thermal conductivity $\lambda$ is also considered in Eq. (\ref{eq_ene}). Note that the thermal conductivity of polymeric liquids is anisotropic under shear flows in general\cite{art:90BB,art:96OP,art:96BC,art:97BCB}, and some experimental studies have reported that the linear stress-thermal relation between the stress tensor and thermal conductivity holds.\cite{art:01VSIGB,art:04SVBBS,art:12SVG,art:13GSV}
However, in the present study, we only consider the isotropic thermal conductivity as the first step because the effect of shear thinning of the viscosity is thought to be more crucial to viscous heating under strong shear flows than that of the anisotropy of the thermal conductivity.
Involving the anisotropic thermal conductivity in the SMD simulation is an important future work. We also assume that the velocity and temperature of the polymeric liquid are the same as those of the plates at the boundaries, i.e., the non-slip and non-temperature-jump boundary conditions.

The effect of viscous heating is estimated using the ratio of the first and second terms in Eq. (\ref{eq_ene}) to be $\sigma_0\dot\Gamma H^2/\lambda \Delta T_0$. Here, $\dot\Gamma$ is the gross shear rate of the system, which is defined by the ratio of the velocity of the upper plate $v_w$ to the width of the gap $H$, $\dot \Gamma=v_w/H$, and $\Delta T_0$ is a characteristic temperature rise for the polymeric liquid. In the present problem, we consider a characteristic temperature necessary to substantially change the viscosity of the polymeric liquid, i.e., $\Delta T_0=|\eta_0/(\partial \eta_0/\partial T_0)|$, where $\eta_0$ is the characteristic viscosity of the polymeric liquid at a temperature of $T_0$. Thus, the Nahme-Griffith number $Na$, defined as 
\begin{equation}\label{eq_nahme-griffith}
Na=\frac{\sigma_0\dot\Gamma H^2}{\lambda|\partial \log(\eta_0)/\partial T_0|^{-1}} 
\end{equation}
represents the effect of viscous heating on the changes in the rheological properties.\cite{book:87BAH,art:08PMM} Usually, in lubrication systems and in high-speed processing operations with polymeric liquids, the Nahme-Griffith number is not negligibly small because of the large viscosity and the small thermal conductivity of the polymeric liquid.\cite{art:08PMM} For example, when a lubrication oil in a gap with a width of 1 $\mu$m is subjected to shear deformation with a strain rate of $1\times 10^6$ $\rm s^{-1}$, the Nahme-Griffith number is estimated to be $Na\gtrsim 0.1$. Thus, the rheological properties of the lubricant in such micro devices must be significantly affected not only by the large velocity gradient but also by the temperature increase caused by local viscous heating. To predict the rheological properties of the polymeric liquid in these systems, one must consider the temperature variation in Eq. (\ref{eq_ene}) coupled with Eq. (\ref{eq_flow}).

\section{Simulation Method}
\begin{figure}[tb]
\begin{center}
\includegraphics[scale=1]{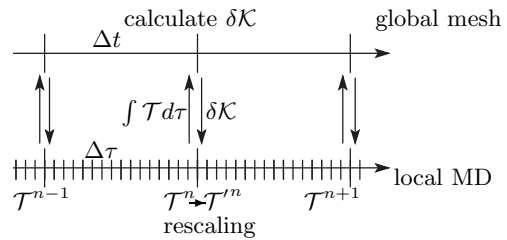}
\caption{%
A schematic for the calculation of the temperature in the SMD method. The upper side represents the progress on the global mesh system, and the lower side represents the progress at each MD cell.
}\label{fig_calc_temp}
\end{center}
\end{figure}

In the present simulation, the gap between the parallel plates is divided into $M$ mesh intervals with a uniform width of $\Delta y=H/M$, and the local velocities are calculated at each mesh node through the typical finite volume scheme shown in Eq. (\ref{eq_flow}). The local shear stresses $\sigma_{xy}(y)$ are calculated from the local shear rates in the MD cells associated with each mesh interval using the NEMD simulation with the SLLOD algorithm. The MD simulations are performed in a time interval $\Delta t$, and the time integrals of the instantaneous shear stresses ${\cal P}_{xy}$ in each MD cell are used to update the local velocities at the next time step in accordance with the macroscopic momentum transport Eq. (\ref{eq_flow}),
\begin{equation}
v_x^n(y)=v_x^{n-1}(y)+\frac{\partial}{\partial y}\int_{(n-1)\Delta t}^{n\Delta t}{\cal P}_{xy}(\tau;\dot \gamma^{n-1}(y))d\tau.
\end{equation}
Here, the superscript $n$ represents the time step number, ${\cal P}_{xy}(\tau;\dot \gamma^{n-1}(y))$ is the instantaneous shear stress in the NEMD simulation with a shear rate of $\dot \gamma^{n-1}(y)$, and $\tau$ is the temporal progress of the NEMD simulation. Note that the time-step size of the MD simulation is different from $\Delta t$. The local viscous heating caused by the shear flow, i.e., the first term of Eq. (\ref{eq_ene}), is calculated in the NEMD simulations without the use of a thermostat algorithm in each MD cell; however, at each time interval $\Delta t$, the instantaneous kinetic energies of the molecules per unit mass $\cal K$ in each MD cell are corrected according to the heat fluxes between neighboring MD cells. Figure \ref{fig_calc_temp} is a schematic of the calculation of the temperature in the SMD method. The instantaneous temperatures {$\cal T$} at each MD cell and their integrals over the duration of each MD run, i.e., $\int_{(n-1)\Delta t}^{n\Delta t}{\cal T}(\tau)d\tau$, are calculated at each MD cell. The heat fluxes between neighboring MD cells $\delta {\cal K}$ are calculated on the global mesh system (depicted on the upper side in Fig. \ref{fig_calc_temp}) as
\begin{equation}
\delta {\cal K}=-\frac{\lambda}{\rho_0} \frac{\partial^2}{\partial y^2}\int_{(n-1)\Delta t}^{n\Delta t}{\cal T}(\tau)d\tau,
\end{equation}
and the instantaneous kinetic energies ${\cal K}$ at each MD cell are corrected by rescaling the molecular velocities according to the corrected temperature ${\cal T'}$ (depicted on the lower side in Fig. \ref{fig_calc_temp}) via
\begin{equation}
{\cal T'}^n={\cal T}^n+\frac{2}{3}\delta {\cal K}.
\end{equation} 
Thus, the MD simulations assigned to each fluid element are synchronized at time intervals of $\Delta t$ to satisfy the macroscopic heat and momentum transport equations (see Fig. \ref{fig_problem}(b)). Note that it is difficult to rewrite Eq. (\ref{eq_ene}) as the time evolution of temperature in general because the internal energy depends on not only the macroscopic variables but also on the conformations of the polymer chains. In the present SMD simulation, the temperature increase caused by local viscous heating is calculated autonomously in the MD simulation and satisfies the macroscopic energy balance of Eq. (\ref{eq_ene}). 

Hereafter, we measure the length, time, temperature and density in units of $\sigma$, $\tau_0=\sqrt{m\sigma^2/\epsilon}$, $\epsilon/k_B$, and $m/\sigma^3$, respectively. Here, $k_B$ is the Boltzmann constant, and $m$ is the mass of the LJ particle. In the following simulations, the density and thermal conductivity of the polymeric liquid are fixed to be $\rho_0=1$ and $\lambda$=150, respectively, and the temperature of the plates and the width of the gap between the plates are fixed to be $T_0=0.2$ and $H=2500$, respectively, whereas the shear stress applied to the upper plate $\sigma_0$ varies. At this number density $\rho_0$ and this temperature $T_0$, the conformation of the bead particles becomes severely jammed and results in complicated rheological properties.\cite{art:11YY,art:02YO}

We have carried out the numerical tests of the present method for various calculation conditions by varying the number of mesh intervals $M$, time interval $\Delta t$, and number of polymer chains in each MD cell $N_p$. The results and a critical analysis of the present method are given in Sec. V. In the present paper, unless otherwise stated, the following calculation parameters are used: The number of mesh intervals $M$ and the time interval $\Delta t$ are $M$=32 and $\Delta t$=1, respectively. A total of 100 polymer chains, i.e., $N_p=100$, of ten beads, i.e., 1000 bead particles, are contained in each MD cell. Thus, the mesh width and side length of the MD cell are $\Delta y$=78.125 and $l_{\rm MD}$=10, respectively. The time-step size in the MD simulation $\Delta \tau$ is set to be $\Delta \tau=0.001$. Thus, MD simulations are performed for 1000 time steps in the time interval $\Delta t$, i.e., $\Delta t=1000\Delta \tau$.

\section{Results}
We performed SMD simulations with various values for the shear stress applied to the plate $\sigma_0$, i.e., $\sigma_0=$0.002, 0.01, 0.03, 0.05, 0.055, 0.06, 0.07, 0.08, and 0.09, and investigated the behaviors of a polymeric liquid at steady state. In the following, we present quantities averaged for a long period of time at steady state, in which the shear stress is spatially uniform and the time derivative of the local internal energy, i.e., the left-hand side of Eq. (\ref{eq_ene}), is negligible.

\begin{figure}[tb]
\begin{center}
\includegraphics*[scale=0.65]{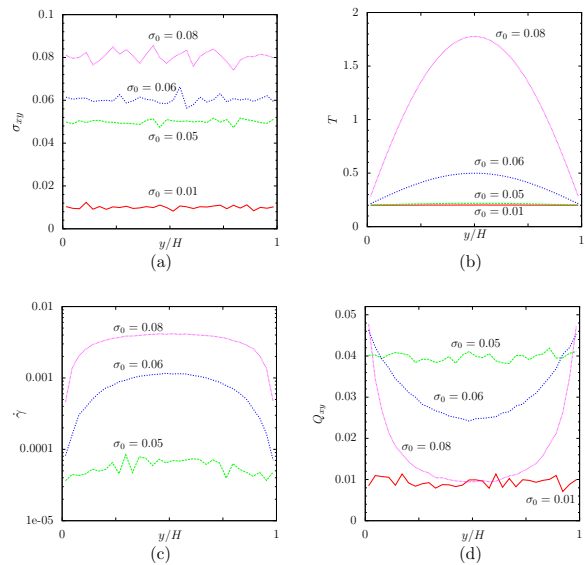}
\end{center}
\caption{%
The spatial variation in the shear stress $\sigma_{xy}(y)$ (a), temperature $T(y)$ (b), shear rate $\dot \gamma(y)$ (c), and $xy$ component of the bond orientation tensor in Eq. (\ref{bond_orientation}) (d) for various values of $\sigma_0$.
}\label{fig_local}%
\end{figure}
Figure \ref{fig_local} shows the spatial variations in local quantities for various values of the shear stress applied to the plate, i.e., for $\sigma_0=$0.01, 0.05, 0.06, and 0.08. The shear stress $\sigma_{xy}$ is found to be spatially uniform, and this fact also indicates that the condition necessary for the polymeric liquid to be at steady state is satisfied in the present simulation. The spatial variation in the temperature $T$ is small when the applied shear stress $\sigma_0$ is smaller than 0.05, i.e., $\sigma_0 \le 0.05$, whereas for $\sigma_0 \ge 0.06$, the spatial variation in $T$ becomes notable. The spatial variation in the shear rate $\dot \gamma$ also increases rapidly when the applied shear stress $\sigma_0$ is larger than 0.05. However, the behaviors of the spatial variations in the temperature and in the shear rate are different. The spatial variation in the shear rate is only enhanced in the vicinity of the plate and is rather moderate, except in the vicinity of the plate, whereas the spatial variation in the temperature is a parabolic curve throughout the region. Note that in Fig. \ref{fig_local}(c), we omit the result for $\sigma_0=$0.01 because the amplitude of the shear rate is very small and occasionally exhibits negative values as a result of the fluctuations; the spatial average of the shear rate for $\sigma_0=$0.01 is $7.2\times 10^{-6}$. In Fig. \ref{fig_local}(d), we show the $xy$ component of the bond orientation tensor $Q_{xy}$, which is defined as
\begin{equation}\label{bond_orientation}
Q_{\alpha\beta}=\frac{1}{N_{\rm p}}\sum_{\rm chain}\frac{1}{N_{\rm b}-1}\sum_{j=1}^{N_{\rm b}-1}\frac{b_{j\alpha}}{b_{\rm min}}\frac{b_{j\beta}}{b_{\rm min}},
\end{equation}
where $N_{\rm p}$ is the number of polymer chains in each MD cell, $N_{\rm b}$ is the number of bead particles in a polymer chain, ${\bm b}_j$ for $1\le j\le N_{\rm b}-1$ is the bond vector between consecutive beads in the same chain, and $b_{\rm min}$ is the distance at which the sum $U_{\rm LJ}(r)+U_{\rm F}(r)$ has a minimum and is calculated to be $b_{\rm min}\simeq 0.97$. The spatial variation in $Q_{xy}$ is also small for small applied shear stresses, i.e., $\sigma_0 \le 0.05$, but this spatial variation becomes notable for $\sigma_0 \ge 0.06$. In the vicinity of the plate, $Q_{xy}$ increases monotonically with increasing applied shear stress $\sigma_0$, whereas in the middle of the plates, $Q_{xy}$ varies non-monotonically with the applied shear stress. This complicated behavior arises from the temperature variation; the conformation of the polymer chains in the vicinity of the plate is mainly influenced by the shear rate because the variation in the temperature from the reference value $T_0$ is small there, but at the middle of the plates, the conformation of the polymer chains is greatly influenced by the temperature because the variation in the temperature from the reference value $T_0$ is large there. The relations among the temperature, shear rate, and conformation of the polymer chains are also discussed in detail later.

\begin{figure}[tb]
\includegraphics[scale=0.6]{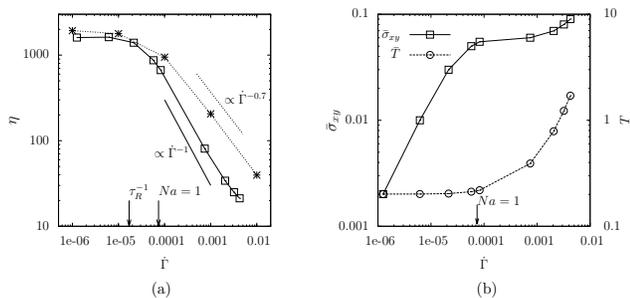}
\caption{%
The apparent viscosity $\eta$, defined as $\eta=\sigma_0/\dot \Gamma$, (a) and the spatial averages of the shear stress and temperature, $\bar \sigma_{xy}$ and $\bar T$, (b) as functions of the gross shear rate $\dot \Gamma$. In figure (a), the asterisks ``$*$'' show the results of the viscosity for a uniform temperature $T=0.2$, i.e., $Na=0$, obtained by the NEMD simulations. The downward arrows in both figures represent the shear rate at which the Nahme-Griffith number $Na$ defined in Eq. (\ref{eq_nahme-griffith}) equals unity.
}\label{fig_gross2}
\end{figure}
\begin{figure}[tb]
\includegraphics[scale=0.7]{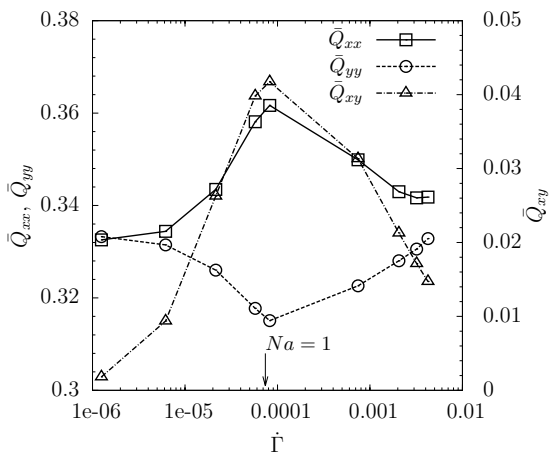}
\caption{%
The spatial average of the bond-orientation tensor of the polymer chains $\bar Q_{\alpha\beta}$ as a function of the gross shear rate $\dot \Gamma$. 
In addition, see the caption in Fig. \ref{fig_gross2}
}\label{fig_qab}
\end{figure}

In the following, we investigate the rheological properties of the system. Hereafter, the spatial average of a quantity $a(y)$ is represented as $\bar a$, i.e., $\bar a=\frac{1}{H}\int_0^Ha(y)dy$. Figure \ref{fig_gross2} shows the apparent viscosity of the system and the spatial averages of the shear stress and temperature as a function of the gross shear rate. The downward arrows in Fig. \ref{fig_gross2} indicate the reference shear rate at which the Nahme-Griffith number $Na$ defined in Eq. (\ref{eq_nahme-griffith}) equals unity, where the reference temperature is calculated from the viscosities of the model polymeric liquid that were obtained by the NEMD simulation at the uniform and constant temperatures of $T=0.2$ and 0.4 in Ref. \onlinecite{art:02YO} as $\Delta T_0=0.15$. In Fig. \ref{fig_gross2}(a), the apparent viscosity is compared to the viscosity of the polymeric liquid with the uniform and constant temperature $T_0$ obtained from the NEMD simulation; this case corresponds to a negligible Nahme-Griffith number, i.e., $Na=0$. Strong shear thinning occurs when the Nahme-Griffith number exceeds unity, i.e., $Na>1$. The slope in the power law, i.e., the index $\nu$ in the approximate relation $\eta(\dot\Gamma)\propto \dot \Gamma^{-\nu}$, is almost unity (but never exceeds unity), i.e., $\nu \simeq 1$, for $\dot \Gamma=1\times 10^{-4}\sim 1\times 10^{-3}$. A discrepancy between the apparent viscosity and that for $Na=0$ is thought to be caused by the fact that the temperature slightly increases in the present simulation even at the smallest gross shear rate (the average temperature $\bar T$ increases by 2.5 percent of the wall temperature $T_0$), and the temporal-spatial fluctuations of the temperature and shear rate are also induced in the present simulation, whereas the temperature and the shear rate are kept constant and uniform in the NEMD simulation. Figure \ref{fig_gross2}(b) shows the spatial averages of the local shear stress and temperature as functions of the gross shear rate. We note that the spatial average of the shear stress $\bar \sigma_{xy}$ coincides with the applied shear stress $\sigma_0$ on the plate because the local shear stress is almost spatially uniform at steady state, as shown in Fig. \ref{fig_local}(b). The average shear stress $\bar \sigma_{xy}$ is observed to increase monotonically with the gross shear rate. The plateau region of the curve corresponds to the strongly shear-thinning regime with the index $\nu\simeq 1$ in Fig. \ref{fig_local} (a). The average temperature increases rapidly with increasing gross shear rate when the Nahme-Griffith number exceeds unity. The rate of increase of the average temperature is lower than that of the average shear stress at small gross shear rates, i.e., $\dot\Gamma \lesssim 1\times 10^{-4}$, but reverses at large gross shear rates, i.e., $\dot\Gamma \gtrsim 1\times 10^{-4}$.

Figure \ref{fig_qab} shows the conformation changes in the polymer chains as a function of the gross shear rate. A non-monotonic dependence of the conformation of the polymer chains on the gross shear rate is observed. A transition occurs at the entrance of the strong shear-thinning regime, with $\nu\simeq 1$ in Fig. \ref{fig_gross2}(a), i.e., $\dot \Gamma\simeq 1\times 10^{-4}$. The polymer chains are stretched in the $x$ direction, and the $xy$ component of the bond-orientation tensor increases as $\dot \Gamma$ for small gross shear rates, i.e., $\dot \Gamma \lesssim 1\times 10^{-4}$. However, for large gross shear rates, i.e., $\dot\Gamma \gtrsim 1\times 10^{-4}$, the alignment of the polymer chains is disturbed, and the conformation of the polymer chains becomes more isotropic as the gross shear rate increases. This transitional behavior is caused by the temperature variation; when the temperature increases sufficiently for the large gross shear rates, the coherent structure becomes disturbed by the thermal motion of the bead particles.

\begin{figure}[tb]
\includegraphics[scale=0.75]{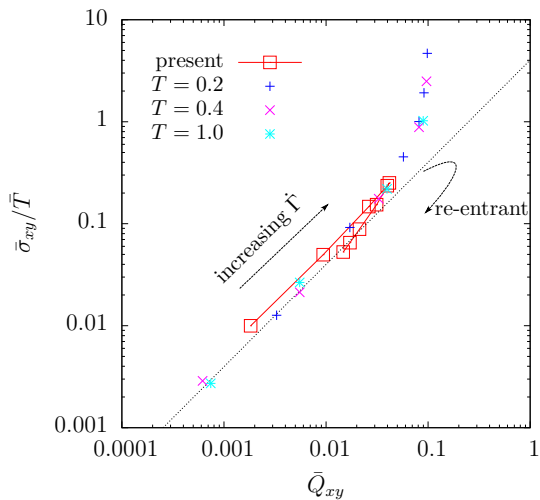}
\caption{%
The stress-optical relation $\bar \sigma_{xy}/\bar T$ vs. $\bar Q_{xy}$.
The squares $\square$ indicate the present results, and the symbols $+$, $\times$, and $*$ indicate the results obtained in Ref. \onlinecite{art:02YO} at uniform temperatures. 
}\label{fig_stress_optic}
\end{figure}
Figure \ref{fig_stress_optic} shows the stress-optical relation for the present problem.\cite{book:83J,book:86DE,book:07S} For the present model polymeric liquid, the NEMD simulations for the isothermal shear flows, i.e., $Na$=0, showed that a universal curve in the stress-optical relation holds in both the linear ($\bar Q_{xy} \lesssim 0.05$) and nonlinear ($\bar Q_{xy} \lesssim 0.05$) regimes.\cite{art:02YO} In the figure, the present results are compared with those for the case $Na=0$. The leftmost square symbol in the present study represents the result for the smallest gross shear rate, i.e., $\dot \Gamma\simeq 1\times 10^{-6}$. It is observed that $\bar \sigma_{xy}/\bar T$ increases with $\bar Q_{xy}$, while the gross shear rate increases up to $\dot\Gamma \lesssim 1\times 10^{-4}$ because the shear stress increases more rapidly with the gross shear rate than does the temperature (see Fig. \ref{fig_gross2}(b)); moreover, $\bar Q_{xy}$ also increases with the gross shear rate in this regime (see Fig. \ref{fig_qab}). However, for large gross shear rates, i.e., $\dot \Gamma \gtrsim 1\times 10^{-4}$, the temperature increases more rapidly than does the shear stress, and the $xy$ component of the bond-orientation tensor decreases with the gross shear rate. Interestingly, the linear stress-optical relation is recovered for shear stresses larger than that for the transitional behavior of the conformation tensor, although the temperature, shear stress, and conformation of the polymer chains exhibit very complicated nonlinear behavior. This reentrant transition of the linear stress-optical relation for large shear stresses can never be reproduced in the NEMD simulations using a thermostat because both the shear stress $\bar \sigma_{xy}$ and the $xy$ component of the bond orientation tensor $\bar Q_{xy}$ monotonically increase with the shear rate at a constant temperature $\bar T$, but $\bar Q_{xy}$ saturates to a limiting value $\bar Q_{xy}\sim 0.1$; therefore, the nonlinear stress-optical relation forms, as shown in Fig. \ref{fig_stress_optic}, for the case $Na=0$.

\section{Critical analysis of the method}
In this section, numerical tests for various calculation conditions are implemented by varying the number of mesh intervals $M$ (i.e., $\Delta x$=$H/M$), time interval $\Delta t$, and number of polymer chains in each MD cell $N_p$, which are given in Table \ref{t1}. The comparisons between the results of the present method and the analytic solutions given by Gavis and Laurence are also presented.\cite{art:68GL}
\begin{table}[htbp]
\caption{Calculation Conditions}\label{t1}
\begin{tabular}{c c c c}
\hline\hline
&$M$&$\Delta t$&$N_p$\\
\hline
C1 &32&1.0&100\\
C2 &64&1.0&100\\
C3 &16&1.0&100\\
C4 &32&0.5&100\\
C5 &32&5.0&100\\
C6 &32&1.0&1000\\
\hline\hline
\end{tabular}
\end{table}
\begin{figure}[tb]
\includegraphics[scale=0.6]{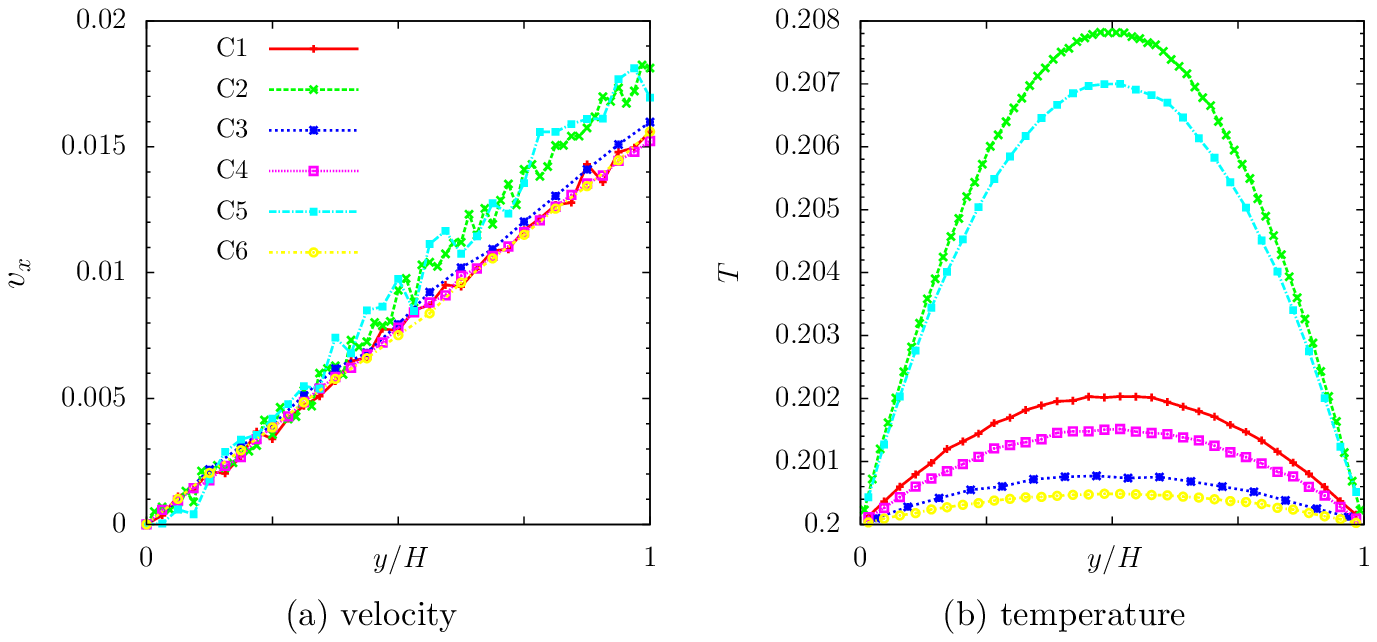}
\caption{%
The comparisons of the velocity (a) and temperature (b) profiles under the various calculation conditions for the applied shear stress $\sigma_0=0.01$.
}\label{fig_compari_pw001}
\end{figure}
\begin{figure}[tb]
\includegraphics[scale=0.6]{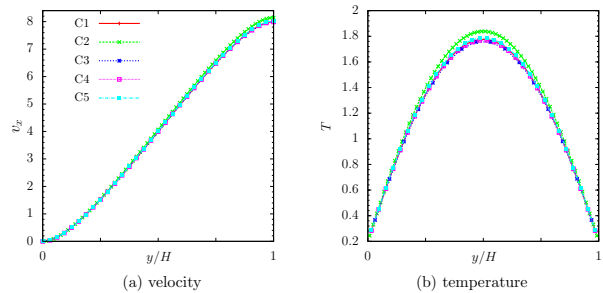}
\caption{%
The comparisons of the velocity (a) and temperature (b) profiles under the various calculation conditions for the applied shear stress $\sigma_0=0.08$. }\label{fig_compari_pw008}
\end{figure}

Figures \ref{fig_compari_pw001} and \ref{fig_compari_pw008} show the comparisons of the velocity and temperature profiles obtained under the different calculation conditions for the applied share stresses $\sigma_0=0.01$ and $\sigma_0=0.08$, respectively. The comparisons between C1, C2, and C3 show the effect of changing the mesh interval $\Delta x$, and the comparisons between C1, C4, and C5 show the effect of changing the time interval $\Delta t$.
Note that the time-step size of the MD simulation $\Delta \tau$ is fixed to be $\Delta \tau=0.001$. Thus, the MD simulations are performed for 500 and 5000 time steps for C4 and C5, respectively, in each time interval $\Delta t$. The comparison between C1 and C6 shows the effect of changing the number of particles in each MD cell, i.e., the effect of the noise intensity arising from the MD simulations.

It is observed in Fig. \ref{fig_compari_pw001} and in \ref{fig_compari_pw008} that for $\sigma_0=0.01$, the velocity profiles for C2 and C5 significantly deviate from those under other calculation conditions, whereas for $\sigma_0=0.08$, the deviations under different calculation conditions are not notable. The deviations of C2 and C5 for $\sigma_0=0.01$ are considered to be a result of the numerical stability condition for the momentum transport equation Eq. (\ref{eq_flow}), which is written as $\Delta t < \Delta x^2/2\eta$ for a Newtonian fluid with a constant viscosity $\eta$. The exact stability condition for the SMD method is unknown because in the SMD simulations, the fluctuations are involved in local stresses, and the local viscosities are also autonomously modulated according to the local flow velocities. In the present numerical tests, only the calculation conditions C2 and C5 do not satisfy the condition $\Delta t < \Delta x^2/2\eta$, where $\eta$ is the proper viscosity obtained under other calculation conditions. A similar situation is also found in the previous study in Ref. \cite{art:10YY}, in which numerical tests for the creep motion of the model polymer melt with a uniform temperature are performed. In the previous study, we found that the multiscale simulation in which the time-step size $\Delta t$ is larger than the viscous diffusion time $\Delta x^2/\eta$, i.e., $\Delta t > \Delta x^2/\eta$, reproduces the velocity profile with a pseudo viscosity that is smaller than the proper viscosity. These facts indicate that the time interval $\Delta t$ must be smaller than the viscous diffusion time $\Delta x^2/\eta$, within which the viscous force propagates for the mesh interval $\Delta x$, to obtain the proper solutions in the SMD simulations. This also explains why the deviations of the solutions between different calculation conditions for $\sigma_0=0.08$ are small because for $\sigma_0=0.08$, the local viscosities decrease as a result of both the shear thinning and the temperature increase so that the relation $\eta \ll \Delta x^2/\Delta t$ is satisfied everywhere, except at the vicinity of the plate, where the temperature is close to the reference temperature. Incidentally, a similar condition is also required for the energy transport equation, but it is usually satisfied as long as that for the momentum transport equation is satisfied because the thermal conductivity is usually small compared to the viscosity (i.e., the Prandtl number is small) for polymeric liquids.

The temperature profiles for $\sigma_0=0.01$ do not coincide with each other, although the absolute differences are small, i.e., at most one percent of the wall temperature $T_0$, while for $\sigma_0=0.08$, the deviations between different calculation conditions are very small. This may be caused by the fluctuations of local shear rates and stresses because the signal-to-noise ratio is smaller for $\sigma_0=0.01$ than for $\sigma_0=0.08$. However, by increasing the particle numbers in each MD cell, the fluctuations due to the noise arising from the MD simulations can be reduced.
\begin{figure}[tb]
\includegraphics[scale=0.55]{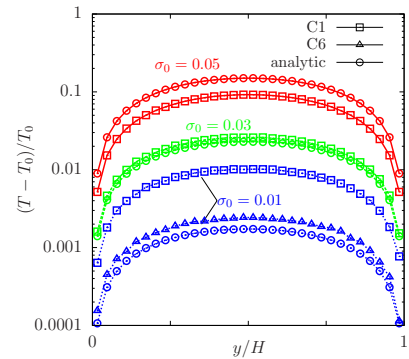}
\caption{%
The comparisons of temperature profiles between the SMD simulations and the analytic solutions for various shear stresses.}\label{fig_compari_analy}
\end{figure}
Figure \ref{fig_compari_analy} shows the comparisons between the SMD simulations and the analytic solutions for different applied shear stresses. Here, the analytic solution is obtained for a Newtonian liquid with an exponential dependence of the viscosity on temperature\cite{art:68GL,art:00BM}, i.e., $\eta(T)=\eta_0\exp[-(T-T_0)/\Delta T_0]$, where $\eta_0$ is a characteristic viscosity at a uniform temperature $T_0$ and $\Delta T_0$ is defined above Eq. (\ref{eq_nahme-griffith}). The analytic solution for the temperature is written as\cite{art:68GL,art:00BM} 
%
%\begin{equation}
\begin{align}
T(\hat y)&=T_0+\Delta T_0\times \nonumber \\
&\log\left\{\left(1+\frac{\tilde {Na}}{8}\right){\rm sech}^2\left[\left({\rm arcsinh}\sqrt{\frac{\tilde {Na}}{8}}\right)(2\hat y-1)\right]\right\},
\end{align}
%\end{equation}
%
where $\hat y=y/H$, and $\tilde {Na}$ is obtained by replacing $\sigma_0$ with $\eta_0\dot \Gamma$ in Eq. (\ref{eq_nahme-griffith}). Here, the characteristic viscosity $\eta_0$ is estimated using the NEMD simulation with a uniform temperature $T_0$ at a shear rate $\dot \gamma=1\times 10^{-6}$ as $\eta_0=1930$.

In Fig. \ref{fig_compari_analy}, it can be observed that the temperature profile for $\sigma_0=0.01$ under the calculation condition C1 deviates considerably from the analytic solution; however, under the calculation condition C6, where the number of polymer chains in the MD cell is increased tenfold, the temperature profile becomes much closer to the analytic solution. Note that for $\sigma_0=0.01$, the effect of the shear thinning of the viscosity is very small, as observed in Fig. \ref{fig_gross2}. Thus, the analytic solution is thought to represent an accurate temperature profile for which the noise effects in the SMD simulation are completely neglected. This fact indicates that the temperature profile obtained by the SMD simulation for small applied shear stresses is greatly affected by the noise even after the long-time average is taken. Thus, to obtain an accurate temperature profile for a small applied shear stress, one needs a large number of molecules in each MD cell, although the cell size $l_{\rm MD}$ should be smaller than the mesh interval $\Delta x$ for an efficient computation. The temperature profile obtained using the SMD simulation for $\sigma_0=0.05$ also deviates from the analytic solution; this is caused by shear thinning, which is not considered for the analytic solution.

In addition to the overall technical issues, there are also concerns about the conceptual issues of the SMD method. The synchronous scheme via the macroscopic transport equations imposes ignoring the molecular transports of constituents across the mesh interval. Thus, the SMD method is not applicable for the dilute polydisperse fluids but is rather designed for dense fluids such as the polymer melt. In the concept of locality of the SMD method, the viscous diffusion and the vortex structures are resolved in the global mesh system. Thus, the mesh interval and the size of the MD cell must be sufficiently small such that the fluid inertia and convection can be ignored at those scales, i.e., the local Reynolds number at the mesh interval $\Delta x$ must be very small. This is also related to the technical aspects of the SMD method because this condition warrants exploiting the SLLOD algorithm and the homogeneous rescaling of the kinetic energies in each MD cell.

\section{Summary}
We have proposed a synchronized molecular dynamics simulation via macroscopic heat and momentum transfer and applied this method to the analysis of the lubrication of a polymeric liquid, coupled with viscous heating. The rheological properties and the conformations of the polymer chains are investigated using a non-dimensional parameter, i.e., the Nahme-Griffith number. The SMD simulation demonstrates that strong shear thinning, which is almost inversely proportional to the shear rate, and the transitional behavior for the conformation of the polymer chains occur with a rapid temperature increase when the Nahme-Griffith number exceeds unity. 
The results show that the linear stress-optical relation holds despite the complicated behaviors of the temperature, shear rate, and conformation of the polymer chains. 

We have also carried out numerical tests under various calculation conditions by varying the number of mesh intervals $M$ (i.e., $\Delta x$=$H/M$), time interval $\Delta t$, and number of polymer chains in each MD cell $N_p$ and found the following critical issues in the implementation of the SMD simulations:
\begin{enumerate}
\item The time interval $\Delta t$ must be smaller than the viscous diffusion time $\Delta x^2/\eta$, within which the viscous force propagates for the mesh interval $\Delta x$; 
otherwise, the SMD simulations reproduce the solutions with a pseudo viscosity that is smaller than the true viscosity.
\item The temperature profile at a small applied shear stress is strongly affected by the noise arising from the local MD cells. To obtain an accurate solution, taking the long-time averages as well as increasing the number of particles in each MD cell is required.
\item Concerning the concept of locality in the SMD method, the flow behaviors involving the viscous dissipation and the vortex structures are resolved in the global mesh system.
Thus, the mesh interval $\Delta x$ and the size of the MD cell $l_{\rm MD}$ must be sufficiently small such that the fluid inertia and convection can be ignored at those scales, i.e., the local Reynolds numbers at those scales must be very small.\\
\end{enumerate} 

The first issue concerns the numerical stability condition for the macroscopic transport equations Eq. (\ref{eq_macro}), although the exact stability condition is unknown for the SMD method. The second issue is the consequence of the comparisons between the SMD simulations and the analytic solutions. For the present problem, an SMD simulation using 10,000 particles in each MD cell can successfully reproduce an accurate temperature profile that is described by the analytic solution. The third issue is concerned with the locality concept of the SMD method, and this also involves the technical aspects of the SMD method because the use of the small local Reynolds number warrants exploiting the SLLOD algorithm and the homogeneous rescaling of the kinetic energies at each time interval $\Delta t$, which are described in Sec. III.

Although these issues must be considered, the SMD simulation has two distinctive advantages over the full MD simulations: First, the SMD simulation can reduce the computational effort (i.e., the number of molecules) compared to that of the full MD simulation by a factor of $(l_{\rm MD}/\Delta x)^d$, where $d$ is the dimension number of the macroscopic transport equation. In the present simulation, the factor is 0.128 because $d$ is one.
However, the extension to the two- and three-dimensional cases is also possible by incorporating the algorithms developed in, for example, Refs. \onlinecite{art:08YY, MT2010,MT2011,MT2012}. Second, almost perfect parallelization efficiency (i.e., using $N$ CPUs in the parallel computation speeds up the calculation compared to using a single CPU by $N$ times) is achieved in a parallel computation with as many CPUs as there are MD cells by assigning each CPU to an MD cell.\cite{art:10YY} This efficiency is obtained because each MD simulation is performed independently in the time interval $\Delta t$. The advantage in parallel computation holds not only for the short-range interaction molecular models but also for any complicated molecular models for which the parallelization of MD simulations is difficult. These advantages enable us to analyze the complicated flow behaviors of complex liquids at the macroscopic scales found in actual engineering and biological systems on the basis of the appropriate molecular model.

\begin{acknowledgements}
This study was financially supported by the Hyogo Science and Technology Association and by the Grant for Basic Science Research Projects from The Sumitomo Foundation.
The computations have been performed using the facilities of the Supercomputer Center at the Institute for Solid State Physics, the University of Tokyo.
\end{acknowledgements}


\begin{thebibliography}{100}
%
%
\bibitem{book:89AT}
M. P. Allen and D. J. Tildesley,
{\it Computer Simulation of Liquids},
(Oxford University Press, New York, 1989).
%
\bibitem{book:08EM}
D. J. Evans and G. Morris,
{\it Statistical mechanics of nonequilibrium liquids},
(Cambridge university press, New York, 2008).
%
\bibitem{book:87BAH}
R. B. Bird, R. C. Armstrong, and O. Hassager,
{\it Dynamics of polymeric liquids} Vol. 1 (John Wiley and Sons, New York, 1987).
%
%\bibitem{book:06H}
%Y. Hori,
%{\it Hydrodynamic lubrication},
%(Springer-Verlag, Tokyo, 2006).
%
\bibitem{art:93LO}
M. Laso and H. C. \"Ottinger,
``Calculation of viscoelastic flow using molecular models: the CONNFFESSIT approach'',
J. Non-Newtonian Fluid Mech. {\bf 47}, 1 (1993).
%
\bibitem{art:95FLO}
K. Feigl, M. Laso, and H. C. \"Ottinger,
``CONNFFESSIT approach for solving a two-dimensional viscoelastic fluid problem'',
Macromolecules {\bf 28}, 3261 (1995).
%
\bibitem{art:97LPO}
M. Laso, M. Picasso, H. C. \"Ottinger,
``2-D time-dependent viscoelastic flow calculations using CONNFFESSIT'',
AIChE J. {\bf 43}, 877 (1997).
%
\bibitem{art:99DEO}
M. Dressler, B. J. Edwards, \"Ottinger,
``Macroscopic thermodynamics of flowing polymeric liquids'',
Rheol. Acta {\bf 38}, 117 (1999).
%
%%
%\bibitem{art:97HHB}
%M. A. Hulsen, A. P. G. van Heel and B. H. A. A. van den Brule:
%%``Simulation of viscoelastic flows using Brownian configuration fields''
%J. Non-Newton. Fluid Mech. {\bf 70} (1997) 79.
%%
%\bibitem{99HHB}
%A. P. G. van Heel, M. A. Hulsen and B. H. A. A. van den Brule:
%% "Simulation of the Doi.Edwards model in complex flow''
%J. Rheol. {\bf 43} (1999) 1239.
%%
%\bibitem{2000PHHB}
%E. A. J. F. Peters, A. P. G. van Heel, M. A. Hulsen, and B. H. A. A. van den Brule:
%% "Generalization of the deformation field method to
%%  simulate advanced reptation models in complex flow''
%J. Rheol. {\bf 44} (2000) 811.
%%
%\bibitem{2001HPB}
%M. A. Hulsen, E. A. J. F. Peters and B. H. A. A. van den Brule:
%% ``A new approach to the deformation Fields method for solving
%%   complex flows using integral constitutive equations
%J. Non-Newton. Fluid Mech. {\bf 98} (2001) 201.
%%
\bibitem{art:03EE}
W. E and B. Engquist,
``The heterogeneous multi-scale methods'',
Comm. Math. Sci. {\bf 1}, 87 (2003).
%
\bibitem{art:05RE}
W. Ren and W. E,
``Heterogeneous multiscale method for the modeling of complex fluids and micro-fluidics'',
J. Compt. Phys. {\bf 204}, 1 (2005).
%
\bibitem{art:11MD}
M. \"Muller and K. C. Daoulas,
``Speeding Up Intrinsically Slow Collective Processes in Particle Simulations by Concurrent Coupling to a Continuum Description'',
Phys. Rev. Lett. {\bf 107}, 227801 (2011).
%
\bibitem{art:13BLR}
M. K. Borg, D. A. Lockerby, J. M. Reese,
``A multiscale method for micro/nano flows of high aspect ratio'',
J. Compt. Phys. {bf 233}, 400 (2013).
%
%\bibitem{art:07EELRV}
%W. E, B. Engquist, X. Li, W. Ren and E. Vanden-Eijnden,
%%``Heterogeneous multiscale methods: a review'',
%Commun. Comput. Phys. {\bf 2}, 367 (2007).
%%
%
\bibitem{art:03KGHKRT}
I. G. Kevrekidis, C. W. Gear, J. M. Hyman, P. G. Kevrekidis, O. Runborg, and C. Theodoropoulos,
``Equation-free, coarse-grained multiscale computation: enabling microscopic simulations to perform system-level analysis'',
Comm. Math. Sci. {\bf 1}, 715 (2003).
%
\bibitem{art:09KS}
I. G. Kevrekidis and G. Samaey,
``Equation-free multiscale computation: algorithms and applications'',
Annu. Rev. Phys. Chem. {\bf 60}, 321 (2009).
%
\bibitem{art:06DFSKK}
S. De, J. Fish, M. S. Shephard, P. Keblinski, and S. K. Kumar,
``Multiscale modeling of polymer rheology'',
Phys. Rev. E {\bf 74}, 030801(R) (2006).
%
\bibitem{art:13D}
S. De,
``Computational study of the propagation of the longitudinal velocity in a polymer melt contained within a cylinder using a scale-bridging method'',
Phys. Rev. E {\bf 88}, 052311 (2013).
%
%\bibitem{art:10KOK}
%D. A. Kessler, E. S. Oran, and C. R. Kaplan,
%%``Towards the development of a multiscale, multiphysics method for the simulation of rarefied gas flows'',
%J. Fluid Mech. {\bf 661}, 262 (2010).
%
%\bibitem{art:12SSJDNA}
%M. Salloum, K. Sargsyan, R. Jones, B. Debusschere, H. N. Najm, and H. Adalsteinsson,
%%``A stochastic multiscale coupling scheme to account for sampling noise in atomistic-to-conntinuum simulations'',
%Multiscale Model. Simul. {\bf 10}, 550 (2012).
%%
%\bibitem{art:13BAGLRV}
%E. Brini, E. A. Algaer, P. Ganguly, C. Li, F. Rodriguez-Ropero and N. F. A. van der Vegt,
%``Systematic coarse-grainin methods for soft matter simulations - a review'',
%Soft Matter {\bf 9}, 2108 (2013).
%
\bibitem{MT2010}
T. Murashima and T. Taniguchi,
``Multiscale Lagrangian Fluid Dynamics Simulation for Polymeric Fluid''
J. Polym. Sci. B {\bf 48}, 886 (2010).
%
\bibitem{MT2011}
T. Murashima and T. Taniguchi,
``Multiscale Simulation of History Dependent Flow in Polymer Melt'',
Europhys. Lett. {\bf 96}, 18002 (2011).
%
\bibitem{MT2012}
T. Murashima and T. Taniguchi,
``Multiscale Simulation of History Dependent Flow in Polymer Melt'',
J. Phys. Soc. Jpn. {\bf 81}, SA013 (2012).
%
\bibitem{art:08YY}
S. Yasuda and R. Yamamoto,
``A model for hybrid simulation of molecular dynamics and computational fluid dynamics'',
Phys. Fluids {\bf 20}, 113101 (2008).
%
\bibitem{art:09YY}
S. Yasuda and R. Yamamoto,
``Rheological properties of polymer melt between rapidly oscillating plates: an application of multiscale modeling'',
Europhys. Lett. {\bf 86}, 18002 (2009).
%
\bibitem{art:10YY}
S. Yasuda and R. Yamamoto,
``Multiscale modeling and simulation for polymer melt flows between prallel plates'',
Phys. Rev. E {\bf 81}, 036308 (2010).
%
\bibitem{art:11YY}
S. Yasuda and R. Yamamoto,
``Dynamic rheology of a supercooled polymer melt in nonuniform oscillating flows between rapidly oscillating plates'',
Phys. Rev. E {\bf 84}, 031501 (2011).
%
\bibitem{art:13MYTY}
T. Murashima, S. Yasuda, T. Taniguchi, and R. Yamamoto,
``Multiscale Modeling for polymeric flow: particle-fluid bridging scale methods'',
J. Phys. Soc. Jpn. {\bf 82}, 012001 (2013).
%
\bibitem{art:99JAT}
A. Jabbarzadeh, J. D. Atkinson, and R. I. Tanner,
``Wall slip in the molecular dynamics simulation of thin films of hexadecane'',
J. Chem. Phys. {\bf 110}, 2612 (1999).
%
\bibitem{art:01KPY}
R. Khare, J. de Pablo, and A. Yethiraj,
``Molecular simulation and continuum mechanics investigation of viscoelastic properties of fluids confined to molecularly thin films'',
J. Chem. Phys. {\bf 114}, 7593 (2001).
%
\bibitem{art:97KPY}
R. Khare, J. de Pablo, and A. Yethiraj,
``Molecular simulation and continuum mechanics study of simple fluids in non-isothermal planar couette flows'',
J. Chem. Phys. {\bf 107}, 2589 (1997).
%
\bibitem{art:10KBC}
B. H. Kim, A. Beskok, and T. Cagin,
``Viscous heating in nanoscale shear driven liquid flows'',
Microfluid Nanofluid {\bf 9}, 31 (2010).
%
\bibitem{art:90KG}
K. Kremer and G. S. Grest,
``Dynamics of entabgled linear polymer melts: A molecular-dynamics simulation,''
J. Chem. Phys. {\bf 92}, 5057 (1990).
%
\bibitem{art:90BB}
B.H.A.A. van den Brule and S.B.G. O'Brien,
``Anisotropic conduction of heat in a flowing polymeric material'',
Rheol Acta {\bf 29}, 580 (1990).
%
\bibitem{art:96OP}
H. C. \"Ottinger and F. Petrillo,
``Kinetic theory and transport phenomena for a dumbbell model under nonisothermal
conditions'',
J. Rheol. {\bf 40}, 857 (1996).
%
\bibitem{art:96BC}
R. B. Bird and C. F. Curtiss,
``Nonisothermal polymeric fluids'',
Rheol. Acta {\bf 35}, 103 (1996).
%
\bibitem{art:97BCB}
R. B. Bird, C. F. Curtiss, and K. J. Beers,
``Polymer contribution to the thermal conductivity
and viscosity in a dilute solution'',
Rheol. Acta {\bf 36}, 269 (1997).
%
\bibitem{art:01VSIGB}
D. C. Venerus, J. D. Schieber, H. Iddir, J. Guzm\'an, and A. Broerman,
``Anisotropic Thermal Diffusivity Measurements in
Deforming Polymers and the Stress-Thermal Rule'',
Int. J. Thermophysics {\bf 22}, 1215 (2001).
%
\bibitem{art:04SVBBS}
J. D. Schieber, D. C. Venerus, K. Bush, V. Balasubramanian, and S. Smoukov,
``Measurement of anisotropic energy transport in flowing polymers by using a holographic technique'',
PNAS {\bf 101}, 13142 (2004).
%
\bibitem{art:12SVG}
J. D. Schieber, D. C. Venerus, and S. Gupta,
``Molecular origins of anisotropy in the thermal conductivity of deformed polymer melts: stress versus orientation contributions''.
Soft Matter {\bf 8}, 11781 (2012).
%
\bibitem{art:13GSV}
S. Gupta, J. D. Schieber, and D. C. Venerus,
``Anisotropic thermal conduction in polymer melts in uniaxial elongation flows'',
J. Rheol. {\bf 57}, 427 (2013).
%
\bibitem{art:08PMM}
C. J. Pipe, T. S. Majmudar, and G. H. McKinley,
``High shear rate viscometry'',
Rheol Acta {\bf 47}, 621 (2008).
%
\bibitem{art:02YO}
R. Yamamoto and A. Onuki,
``Dynamics and rheology of a supercooled polymer melt in shear flow,''
J. Chem. Phys. {\bf 117}, 2359 (2002).
%
\bibitem{book:83J}
H. Janeschitz-Kriegl, 
{\it Polymer Melt Rheology and Flow Birefringence}
(Springer, Berlin, 1983).
%
\bibitem{book:86DE}
M. Doi and S. F. Edwards,
{\it The theory of polymer dynamics}
(Clarendon, Oxford, 1986).
%
\bibitem{book:07S}
G. Strobl,
{\it The Physics of polymers}
(Springer, Heidelberg, 2007). 
%
\bibitem{art:68GL}
J. Gavis and R. L. Laurence,
``viscous heating in plane and circular flow between moving surfaces'',
I\&EC Fundamentals {\bf 7}, 232 (1968).
%
\bibitem{art:00BM}
L.E. Becker and G.H. McKinley,
``The stability of viscoelastic creeping plane shear flows with viscous heating'',
J. Non-Newton. Fluid Mech. {\bf 92}, 109 (2000).
%
\end{thebibliography}
\end{document}